\documentclass[cits]{PoS}
\usepackage{amssymb, amsthm, amsmath, amsfonts}
\usepackage{fontenc,times,mathptmx}
\usepackage{graphicx,graphics}
\newcommand*{\Tr}[0]{{\rm Tr}}
\newcommand{\field}[1]{\mathbb{#1}} % requires amsfonts

\title{Phase diagram of adjoint QCD at weak coupling and finite volume}

\ShortTitle{Volume dependence of massive adjoint QCD at weak coupling}

\author{Timothy J. Hollowood and \speaker{Joyce C. Myers}\thanks{JCM would like to thank the Royal Society of London for providing the opportunity to present this research.}\\
        Swansea University, Physics Department, 
Vivian Tower, Singleton Park, Swansea SA2 8PP, UK\\
        E-mail: \email{t.hollowood@swan.ac.uk},\,\,\email{j.c.myers@swan.ac.uk}}
        
\abstract{The phase diagram of $SU(N)$ gauge theories with fermions in an arbitrary representation $R$ can be calculated on finite volume manifolds such as $S^1 \times S^3$. When $S^3$ is small a perturbative analysis is possible and the weak-coupling analogue of the pure Yang-Mills theory confinement-deconfinement transition is accessible in the large $N$ limit. We calculate the large $N$ phase diagram of adjoint QCD [$SU(N)$ gauge theory with adjoint fermions] where periodic boundary conditions are applied to fermions on $S^1$ such that the confined phase is favored for light enough adjoint fermion mass $m$. We calculate the value of $m R_{S^3}$ below which the confined phase is favored for all $L_{S^1}/R_{S^3}$ and discuss the implications for large $N$ volume reduction. We calculate also the phase diagram for $N = 3$ and compare with recent lattice results.}

\FullConference{The XXVII International Symposium on Lattice Field Theory - LAT2009\\
		 July 26-31 2009\\
		 Peking University, Beijing, China}

\begin{document}

\section{Introduction}

The calculation of observable quantities in the confined phase of QCD and many QCD-like theories is complicated by the fact that the coupling strength is large. Lattice simulations are the dominant technique for performing calculations in the confining, strong coupling limit, however, they can be computationally demanding since large lattice sizes are often required. In this proceedings we discuss aspects of two shortcuts that can be taken together (or separately) to obtain quick qualitative results from QCD and QCD-like theories. It is based on our recent paper \cite{Hollowood:2009sy}.

The first technique involves the use of two large $N$ equivalences. One, called orientifold planar equivalence, is the large $N$ equivalence between $SU(N)$ [or $U(N)$] gauge theory with adjoint fermions (adjoint QCD), and $SU(N)$ gauge theory with symmetric or antisymmetric representation fermions [QCD(AS/S)] \cite{Armoni:2003gp}. This is true as long as charge conjugation symmetry is not broken in QCD(AS/S) \cite{Unsal:2006pj}, and for only the bosonic subsector of adjoint QCD. In the case of $N = 3$, QCD(AS) is equivalent to QCD, so the large $N$ limit of QCD(AS) is also a large $N$ limit of QCD. The second equivalence is the large $N$ equivalence between different volumes of adjoint QCD \cite{Kovtun:2007py}. This holds as long as the theory is in the confined phase in both volumes. It is a generalization of large $N$ Eguchi-Kawai volume reduction originally proposed for Yang-Mills theory \cite{Eguchi:1982nm}, with the important exception that in adjoint QCD it is conjectured that the center symmetry does not break in the small volume limit \cite{Kovtun:2007py}, as it does in Yang-Mills theory \cite{Kiskis:2003rd}. This is supported by the lattice simulations in \cite{Bringoltz:2009kb}. The overall idea is to use these equivalences to study the a large $N$, large volume limit of QCD by means of simulations of adjoint QCD in small volumes.

The second technique involves implementation of QCD or QCD-like theories analytically on a finite volume, with small enough spatial volume that perturbative analysis becomes valid. Specifically, we implement adjoint QCD on the sphere, $S^1 \times S^3$, following the technique in \cite{Aharony:2003sx}. Perturbative analysis is valid when the size of the compact space is small compared with the strong coupling scale, $\mathrm{min} \left[ R_{S^1}, R_{S^3} \right] \ll \Lambda_{QCD}^{-1}$. In addition, when we take $R_{S^3}$ to be small, then the weak-coupling analogue of the confining-deconfining transition of pure Yang-Mills theory is accessible at large $N$ \cite{Aharony:2003sx}. There is a trade-off which makes this possible. In order to have true phase transitions it is necessary to have an infinite number of degrees of freedom. This can be achieved by taking the volume to be large, or taking the number of colors, $N$, to be large.

It is first in the weak-coupling large $N$, small volume limit that we work. However, it is also possible to uncover the phase diagram at smaller $N$, at small volumes, with the caveat that the transitions are smoothed out and it would be necessary to take the infinite volume limit to make them true, sharp, transitions. It is particularly interesting to compare the $N = 3$ results for the phase diagram of adjoint QCD on $S^1 \times S^3$ with the  lattice results of Cossu and D'Elia \cite{Cossu:2009sq}. Adjoint QCD, with fermions of finite mass $m$, has a rich phase diagram with not only confining, and deconfining phases, but partially confining phases as well, so it serves well to show the amount of agreement possible in comparisons between these two methods. The phase diagram on $S^1 \times S^3$ also shows that the confining region, for which large $N$ volume independence holds, persists for all $L_{S^1} / R_{S^3}$ when $m R_{S^3}$ is below some critical value.

\section{One-loop effective action}

Following \cite{Hollowood:2009sy,Aharony:2003sx,Hollowood:2006xb,Myers:2009df}, one can derived the one-loop effective action on $S^1 \times S^3$ for $SU(N)$ gauge theory with $N_f^D$ Dirac flavors of fermions in the representation ${\cal R}$, and with mass $m$ and chemical potential $\mu$. The only zero mode, $\alpha$, is given by the average of the temporal gauge field $A_0$ over the volume of the sphere:

\[
\alpha \equiv \frac{1}{Vol(S^1 \times S^3)} \int_{S^1 \times S^3} {\mathrm d} \tau \,\, {\mathrm d}^3 x \,\, A_0 ({\bf x}) .
\]

\noindent In terms of the order parameter for the confinement-deconfinement transitions, the Polyakov loop $P = e^{L \alpha} = \text{diag}\{e^{i \theta_1}, ... , e^{i \theta_N}\}$, the effective action is (neglecting the Casimir term)

\begin{equation}
\begin{aligned}
S(P) = &\sum_{n=1}^{\infty} \frac{1}{n} \left( 1 - z_b (n L/R) \right) \Tr_{A} P^n\\
&+ 2 \sum_{n=1}^{\infty} \frac{(\pm 1)^n}{n} N_f^D z_f (n L/R, m R) \left[ e^{n L \mu} \Tr_{{\cal R}} (P^{\dagger n}) + e^{-n L \mu} \Tr_{{\cal R}} (P^n) \right] ,
\end{aligned}
\end{equation}

\noindent where $L$ is the length of $S^1$, and $R$ is the radius of $S^3$. $\Tr_A$ indicates a trace in the adjoint representation. The top sign in $(\pm 1)^n$ corresponds to the case of periodic boundary conditions on $S^1$ for fermions and the bottom sign is for antiperiodic (thermal) boundary conditions. The first term, $\sum_{n=1}^{\infty} \frac{1}{n} \Tr_A P^n$, results from the Jacobian in the partition function for converting from integrals over $SU(N)$ gauge fields to integrals over the Polyakov loop eigenvalue angles. The second and third terms, containing $z_b$ and $z_f$, are the bosonic and fermionic contributions, respectively. These are determined from the energies $\varepsilon_l$ and degeneracies $d_l$ resulting from the action of the laplacian on vector and fermion fields on $S^1 \times S^3$. They are given by \cite{Hollowood:2009sy,Aharony:2003sx}

\begin{equation}
\begin{aligned}
z_b(nL/R) &= \sum_{l=0}^{\infty} d_l^{(v,T)} e^{-n L \varepsilon_l^{(v,T)}} = \sum_{l=0}^{\infty} 2 l (l+2) e^{-n L (l+1)/R}\\
z_f (n L/R, m R) &= \sum_{l=0}^{\infty} d_l^{(f)} e^{-n L \sqrt{\varepsilon_l^{(f) 2}+m^2}} = 2 \sum_{l=1}^{\infty} l(l+1) e^{-n L \sqrt{(l+1/2)^2+m^2 R^2}/R} ,
\end{aligned}
\end{equation}

\noindent where $(v,T)$ indicates that the contribution to the bosonic term is from the action of the laplacian on transverse vector fields.

In what follows we concentrate on adjoint QCD, with zero chemical potential \footnote{It is also interesting to study QCD at finite chemical potential on $S^1 \times S^3$, which is currently under investigation \cite{Hollowood:2009de}.}. The one-loop effective action for adjoint QCD with $N_f$ Majorana fermion flavors with mass $m$, and chemical potential $\mu = 0$ simplifies to

\begin{equation}
S(P) = \sum_{n=1}^{\infty} \frac{1}{n} \left( 1 - z_v (n L/R) + N_f z_f (n L/R, m R) \right) \sum_{i, j=1}^{N} \cos [n (\theta_i - \theta_j)] ,
\end{equation}

\noindent where the application of periodic boundary conditions on the adjoint fermions makes the fermion contribution to the effective action positive, such that this term favors the confined phase, where $\Tr P^n = 0$. Periodic boundary conditions on fermions are required for large $N$ volume independence in adjoint QCD \cite{Kovtun:2007py}.

\section{Obtaining the phase diagram}

To calculate observables it is necessary to consider the partition function for adjoint QCD,

\begin{equation}
\begin{aligned}
Z (L/R) &= \int \left[ {\mathrm d} \theta \right] \text{exp}\{ - \sum_{n=1}^{\infty} \frac{1}{n} \left( 1 - z_v (n L/R) + N_f z_f (n L/R, m R) \right) \left| \Tr P^n \right|^2 \} ,
\end{aligned}
\end{equation}

\noindent where the integrals over the $SU(N)$ gauge fields have been converted into integrals over the Polyakov loop eigenvalue angles with the Jacobian, $\text{exp}\{-\sum_{n=1}^{\infty} \frac{1}{n} \left| \Tr P^n\right|^2 \}$. Notice that in the large $R$ limit, i.e., on $S^1 \times {\field R}^3$, the Jacobian term is subdominant and does not contribute. In the large $R$ limit we can use the saddle point approximation to evaluate the partition function by considering the effective action as the (finite) effective potential multiplied by the (infinite) 4-volume. We can also use the saddle point approximation in the large $N$ limit by normalizing the Polyakov loop appropriately, $\rho_n \equiv \frac{1}{N} \Tr P^n$. The partition function is then

\begin{equation}
\begin{aligned}
Z (L/R) &= \int \left[ {\mathrm d} \theta \right] \text{exp}\{ - N^2 \sum_{n=1}^{\infty} \frac{1}{n} f_n \left| \rho_n \right|^2 \} ,
\end{aligned}
\end{equation}

\noindent where we defined $f_n \equiv 1 - z_v (n L/R) + N_f z_f (n L/R, m R)$. Note that on $S^1 \times S^3$ the saddle point approximation is only strictly valid when $N$ is large. However, for smaller values of $N$, even as small as $N = 2, 3$, we can show that the saddle point approximation still picks out the most favoured configuration, however, nearby configurations also contribute, causing the transitions to be smoothed out. The phase diagram can be obtained by using the saddle point approximation to determine the preferred configurations, in the limit of small $N$, but it is very important to check that the global minima of $S$ don't have closely competing local minima and to compare the results for $\Tr P$ against plots of $e^{-S} \Tr P / Z$ as a function of the full configuration space, as discussed in \cite{Hollowood:2009sy}.

In the large $N$ limit it is helpful to consider a distribution of the Polyakov loop eigenvalue angles. To this end we define the distribution

\begin{equation}
\rho (\theta) \equiv \frac{1}{N} \sum_{i=1}^{N} \delta ( \theta - \theta_i )
\end{equation}

\noindent such that $\rho_n = \int e^{i n \theta} \rho (\theta) {\mathrm d} \theta = \frac{1}{N} \Tr (P^n)$. This allows us to Fourier analyze the distribution

\begin{equation}
\rho (\theta) = \frac{1}{2 \pi} \sum_{n = - \infty}^{\infty} \rho_{-n} e^{i n \theta} ,
\end{equation}

\noindent where $\rho_{-n} = \rho_n^*$ and $\rho_0 = 1$.

The phase diagram is obtained by minimizing the action $S = N^2 \sum_{n=1}^{\infty} \frac{1}{n} f_n \left| \rho_n \right|^2$ for various dimensionless quantities $m L$, $m R$, and $L / R$, to determine the preferred values of the $\rho_n$. The details of obtaining the phase diagram can be found in \cite{Hollowood:2009sy}, but the basic idea is that the sign of the $f_n$ determine the whether it is preferable to maximize or minimize the $\left| \rho_n \right|$. If all the $f_n$ are positive then the preferred phase determined by the minimum of the action is obtained when all the $\rho_n = 0$. This corresponds to the confined phase. However, if one of the $f_k$ becomes negative, then we see a transition, where the corresponding $\rho_k \ne 0$ is preferred. For negative $f_k$ the equations of possible negative actions in the space of all the $\left| \rho_n \right|$ take the form of two-sheeted hyperboloids pointing in the direction of $\left| \rho_k \right|$. The minimum action configuration is defined by the point where the boundary of the allowed configuration space, $\rho (\theta) \ge 0$, is tangent to one of these hyperboloids.

If one plots $\rho (\theta)$ in the region where one of the $f_k < 0$, it becomes clear that this distribution of Polyakov loop angles around the circle has $k$ gaps, and therefore we say the system is in the $k$-gap phase, which is generally defined by

\begin{equation}
\begin{aligned}
&\rho_k = \frac{1}{N} \Tr P^k \ne 0,\\
&\rho_l = \frac{1}{N} \Tr P^l = 0, \,\,\text{for}\,\, \mathrm{mod}[l,k] \ne 0 .
\end{aligned}
\end{equation}

\noindent Of course the 1-gap phase is the deconfined phase with $\left| \rho_1 \right|$ maximized.

\section{Large $N$ results}

\begin{figure}[t]
  \hfill
  \begin{minipage}[t]{.45\textwidth}
    \begin{center}
      \includegraphics[width=0.95\textwidth]{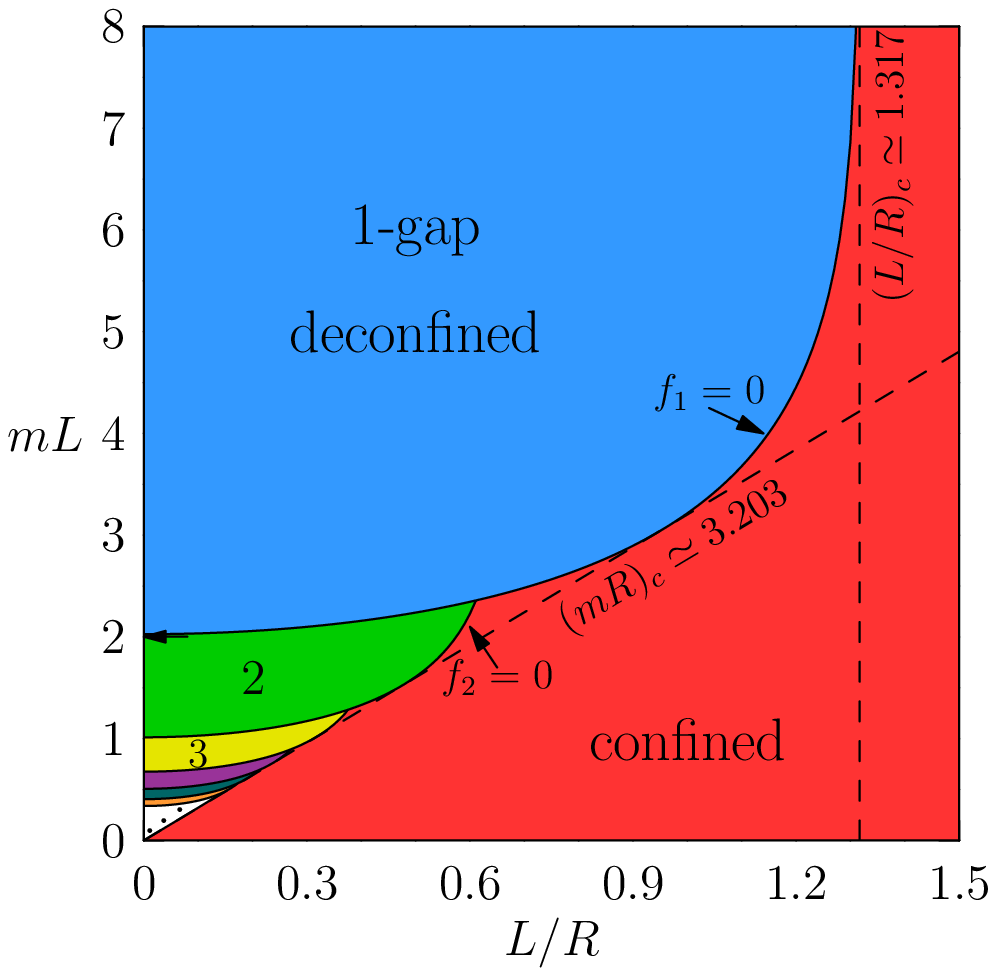}
    \end{center}
  \end{minipage}
  \hfill
  \begin{minipage}[t]{.45\textwidth}
    \begin{center}
\includegraphics[width=0.95\textwidth]{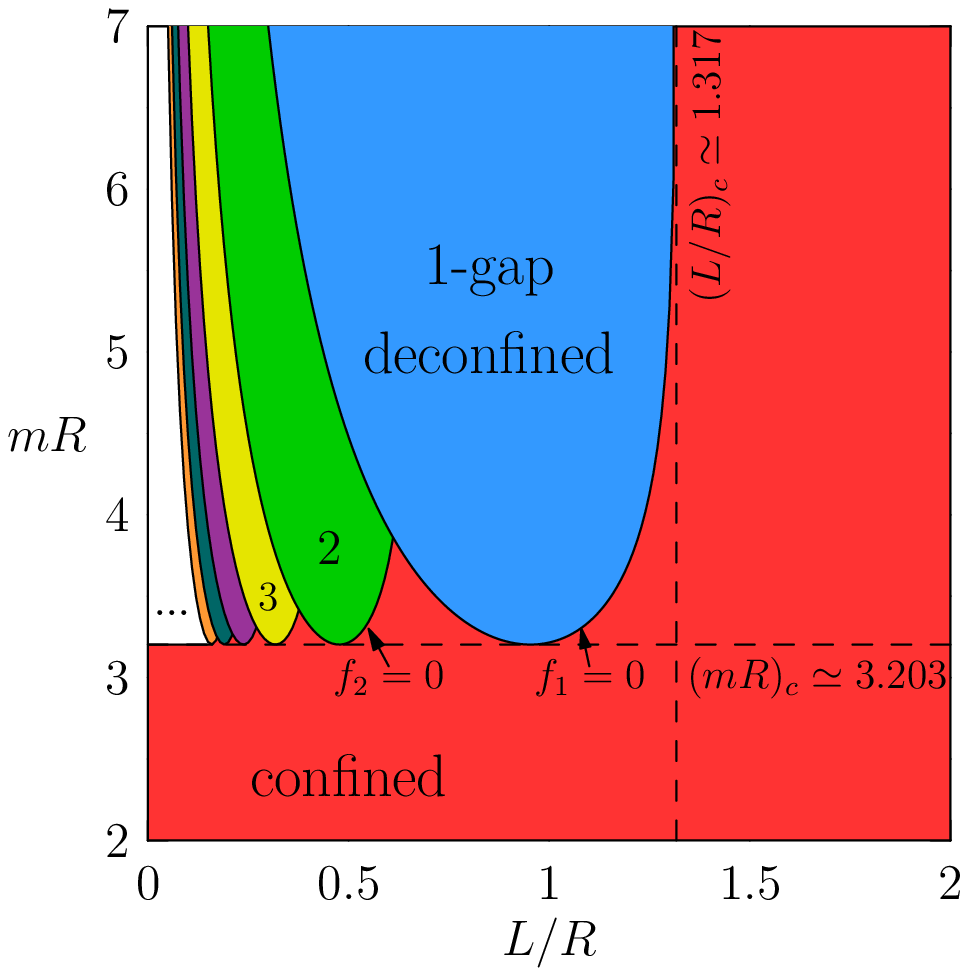}
    \end{center}
  \end{minipage}
  \hfill
  \caption{Phase diagram of adjoint QCD at large $N$ for $N_f = 2$ in the (Left) $(L/R, m L)$ plane and (Right) $(L/R, m R)$ plane.} 
  \label{largeN}
\end{figure}

To show how the large $N$ phase diagram of adjoint QCD depends on the fermion mass and the volume we plot the phase diagram as a function of the dimensionless quantities $L/R$, $m L$, and $m R$. This is shown for $N_f = 2$ in Figure \ref{largeN} in the $(L/R, m L)$ plane (Left) and $(L/R, m R)$ plane (Right). For $N_f = 1$ the results are similar except that only the confined and deconfined phases are observed \cite{Hollowood:2009sy}. For $N_f \ge 2$ we find infinite possible phases. However, the confined phase persists for all $L / R$ when $m R$ is below a certain critical value which increases with $N_f$. Above $(m R)_{c}$ the gapped phases persist for all $m L$ if $L/R \rightarrow 0$, which is the limit of $S^1 \times {\field R}^3$. The confinement-deconfinement transition of the pure $SU(N)$ Yang-Mills theory is indicated by the $(L/R)_{c} = 1.317$ line \cite{Aharony:2003sx}.

While the $f_n = 0$ curves are approximate locations of the transitions, the actual transition lines only very slightly differ, for example, on $S^1 \times {\field R}^3$ the transition between the $1$ and $2$-gap phases occurs for $m L \simeq 2.020$ \cite{Myers:2009df} as indicated by the arrow pointing to the $m L$-axis in Figure \ref{largeN} (Left), whereas the $m L$ asymptote of the $f_1 = 0$ curve is given by $m L \simeq 2.027$.

\section{Finite $N$ results and comparison with lattice data}

\begin{figure}[t]
  \hfill
  \begin{minipage}[t]{.48\textwidth}
    \begin{center}  
      \includegraphics[width=0.95\textwidth]{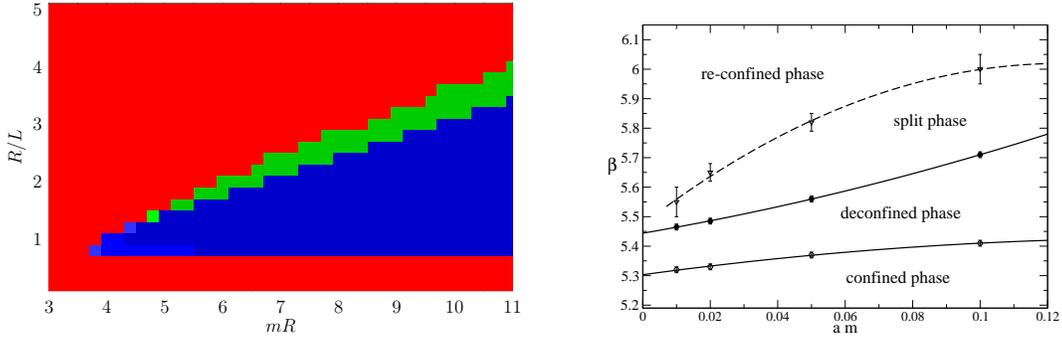}
    \end{center}
  \end{minipage}
  \hfill
  \begin{minipage}[t]{.42\textwidth}
    \begin{center}
\includegraphics[width=0.95\textwidth]{phdiagram.eps}
    \end{center}
  \end{minipage}
  \hfill
\caption{\small QCD(Adj) for $N = 3$, $N_f  = 4$: (Left) ($m R$, $L / R$). $L = 2 \pi R_{S^1}$. Only the confined phase persists for $m R \lesssim 3.6$; (Right) Results from lattice simulations of Cossu and D'Elia \cite{Cossu:2009sq} on a $L_c \times 16^3$ lattice. $\beta$ is related to the inverse coupling $\beta = 2 N / g^2$.} 
  \label{latt_comp}
\end{figure}

\begin{figure}[t]
  \hfill
  \begin{minipage}[t]{.48\textwidth}
    \begin{center}  
      \includegraphics[width=0.95\textwidth]{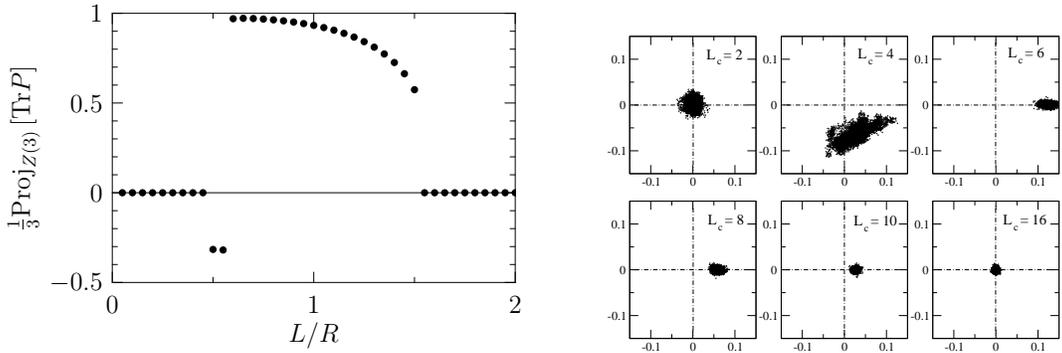}
    \end{center}
  \end{minipage}
  \hfill
  \begin{minipage}[t]{.42\textwidth}
    \begin{center}
\includegraphics[width=0.95\textwidth]{UVfixed.eps}
    \end{center}
  \end{minipage}
  \hfill
  \caption{$\Tr P$ as a function of $L_{S^1}$ for fixed $m$: (Left) $\frac{1}{3}\mathrm{Proj}_{Z(3)} \Tr P$ as a function of $L / R$ on $S^1 \times S^3$ for $m R = 6$. (Right) Histograms of $\mathrm{Im}[\Tr P]$ vs. $\mathrm{Re}[\Tr P]$ for increasing $L_c$ on a $L_c \times 16^3$ lattice \cite{Cossu:2009sq}, with $m a = 0.10$.}
  \label{latt_trp}
\end{figure}
\vspace{-2mm}

In \cite{Cossu:2009sq}, Cossu and D'Elia compute the phase diagram of adjoint QCD using lattice simulations for $N = 3$ and $N_f = 4$ (i.e., $N_f^D = 2$), their results are shown in Figure \ref{latt_comp} (Right). To compare with their results we also calculated the phase diagram for $N = 3$, $N_f = 4$ on $S^1 \times S^3$, by using the saddle point approximation and minimizing the effective action with respect to the Polyakov loop eigenvalue angles, as shown in Figure \ref{latt_comp} (Left). The effective action was additionally plotted in the $\theta_1 \theta_2$-plane to show that the minima determined from the saddle point approximation are well-defined and other local minima compete only minimally. The benefit of using the saddle-point approximation for the $S^1 \times S^3$ results is that we obtain the Polyakov loop eigenvalue angles and can classify the phases accordingly.

Perhaps a better comparison is shown in Figure \ref{latt_trp} where we fix the mass and compare the Polyakov loop as a function of the temporal extent $L$ on $S^1 \times S^3$ with the lattice results in \cite{Cossu:2009sq}. The agreement is good considering that the phases appear in the same order as $L/R$ is increased and that the results even show a similar decline in the magnitude of $\left| \Tr P \right|$ at $L/R$ is increased within the deconfined phase. Of course, using the saddle point approximation for the $S^1 \times S^3$ results causes the transitions to appear sharper than they actually are in a finite volume, and the magnitude $\left| \Tr P \right|$ is actually a bit less, so the agreement is actually even a bit better than what is shown in Figure \ref{latt_trp}.

\section{Conclusions}

The implication of these results for volume independence at large $N$ in adjoint QCD is that the confined phase persists for all $L/R$ when $m R$ is below a certain critical value that increases with $N_f$. This could be checked in lattice simulations. Additionally, the $N = 3$ phase diagram from $S^1 \times S^3$ compares well with the lattice results of \cite{Cossu:2009sq}, with the exception that for adjoint QCD on $S^1 \times S^3$ the confined phase persists for all $L/R$ for small enough $m R$, whereas from the lattice results this is ambiguous. Additionally, the technique used for obtaining the phase diagram on $S^1 \times S^3$ for adjoint QCD could also be used to study QCD and other QCD-like theories.

\acknowledgments{JCM thanks the Royal Society of London for the opportunity to present this work at the Lattice 2009 conference in Beijing. We also thank Guido Cossu and Massimo D'Elia for discussing their related results and allowing us to present their figures in our comparisons with lattice data.}

\end{document}